\documentclass[conference]{IEEEtran}

\IEEEoverridecommandlockouts

\usepackage{cite}
\usepackage{amsmath,amssymb,amsfonts,amsmath}
\usepackage{algorithmic}
\usepackage{graphicx}
\usepackage{tikz}
\usepackage{tikz-3dplot}

\usepackage{hyperref}
\usetikzlibrary{arrows}
\usetikzlibrary{positioning}   
\usepackage{lipsum}
\usetikzlibrary{decorations.markings}
\usetikzlibrary{patterns}
\usetikzlibrary{fit}

\usepackage{textcomp}
\usepackage{xcolor}

\ifCLASSOPTIONcompsoc
    \usepackage[caption=false, font=normalsize, labelfont=sf, textfont=sf]{subfig}
\else
\usepackage[caption=false, font=footnotesize]{subfig}
\fi

\def\BibTeX{{\rm B\kern-.05em{\sc i\kern-.025em b}\kern-.08em
    T\kern-.1667em\lower.7ex\hbox{E}\kern-.125emX}}
\usepackage{mathtools}

\usepackage{amsthm}
\usepackage{amssymb}
\usepackage[utf8]{inputenc}
\usepackage{algorithm2e}
\usepackage{amsmath}

\definecolor{darkcerulean}{rgb}{0.0, 0.0, 0.55}
\definecolor{darkcerulean}{rgb}{0.03, 0.27, 0.49}
\definecolor{darkgreen}{rgb}{0.0, 0.2, 0.13}
\definecolor{applegreen}{rgb}{0.55, 0.71, 0.0}
\definecolor{ashgrey}{rgb}{0.7, 0.75, 0.71}
\definecolor{aurometalsaurus}{rgb}{0.43, 0.5, 0.5}
\definecolor{darkgray}{rgb}{0.66, 0.66, 0.66}

\begin{document}
\tikzset{->-/.style={decoration={
			markings,
			mark=at position #1 with {\arrow{>}}},postaction={decorate}}}

\title{Location- and Orientation-aware Millimeter Wave Beam Selection for Multi-Panel Antenna Devices

\thanks{This work is supported by the Danish Council for Independent Research, grant no. DFF 8022-00371B.}
}

\author{\IEEEauthorblockN{Sajad~Rezaie\IEEEauthorrefmark{1},~João~Morais\IEEEauthorrefmark{2},~Elisabeth~de~Carvalho\IEEEauthorrefmark{1},~Ahmed~Alkhateeb\IEEEauthorrefmark{2},~and~Carles~Navarro~Manch\'on\IEEEauthorrefmark{1}} \IEEEauthorblockA{\IEEEauthorrefmark{1}\textit{Department of Electronic Systems, Aalborg University, Aalborg, Denmark}} \IEEEauthorblockA{\IEEEauthorrefmark{2}\textit{School of Electrical, Computer and Energy Engineering, Arizona State University, Tempe, AZ, USA}}
Email: sre@es.aau.dk, joao@asu.edu, edc@es.aau.dk, alkhateeb@asu.edu, cnm@es.aau.dk}


\maketitle

\begin{abstract}
While initial beam alignment (BA) in millimeter-wave networks has been thoroughly investigated, most research assumes a simplified terminal model based on uniform linear/planar arrays with isotropic antennas. Devices with non-isotropic antenna elements need multiple panels to provide good spherical coverage, and exhaustive search over all beams of all the panels leads to unacceptable overhead. This paper proposes a location- and orientation-aware solution that manages the initial BA for multi-panel devices. We present three different neural network structures that provide efficient BA with a wide range of training dataset sizes, complexity, and feedback message sizes. Our proposed methods outperform the generalized inverse fingerprinting and hierarchical panel-beam selection methods for two considered edge and edge-face antenna placement designs.
\end{abstract}


\section{Introduction} \label{Intro}
Directional beamforming by employing large antenna arrays is the most common way to compensate for the higher propagation and penetration loss at the millimeter-wave (mmWave) and sub-terahertz bands. Codebook-based analog or hybrid analog-digital beamforming are popular solutions for arrays with a number of radio-frequency (RF) chains significantly smaller than the number of antenna elements \cite{giordani_toward_2020}. Finding the optimal beam pair at access point (AP) and  user terminal (UT) using exhaustive search yields unacceptable latency due to the large beam space; hierarchical beam search (HBS), on the other hand, suffers from low signal-to-noise ratio (SNR) at the first search stages. Data driven based methods like inverse fingerprinting (IFP) and machine learning (ML) approaches use prior knowledge of the environment to reduce the search space by proposing a beam candidate list \cite{va_inverse_2018, rezaie_location-_2020}. Besides user location and orientation information, LIDAR, RADAR, and visual images are helpful context information (CI) for beam selection and blockage prediction \cite{xu_3d_2020}. However, most of these methods consider isotropic antenna elements at transceivers.

In practice, antenna arrays at mmWave transceivers are implemented using non-isotropic antenna elements, such as patch antennas, leading to directional coverage. To provide full spherical  coverage, multi-panel antenna designs are often used in UTs  \cite{raghavan_antenna_2019}. 
Such multi-panel designs should be taken into account in both the definition of codebooks  and the design of beam alignment (BA) algorithms. As hand blockage can impact the coverage of a UT, a data-driven approach is proposed in \cite{mo_beam_2019} to generate a non-directional beamforming codebook considering the effects of hand blockage on the antenna radiation pattern. A grip-aware analog beam codebook adaptation provides better spherical coverage over the grip-agnostic scheme by finding a codebook for each hand-gripping mode \cite{alammouri_hand_2019}. Authors in \cite{shih_fast_2021} proposed a beam switching approach to detect hand blockage by utilizing power detectors at all the panels, and in case of blockage, use beams on other panels. However, they focus on beam adjustment after successful initial access. UT panels can be connected to a single or multiple RF chains, possibly with constraints such as each RF chain only being connected to a panel in partially connected hybrid beamforming architectures \cite{song_fully-partially-connected_2020}. Incorporating the limitations and features of multi-panel antenna design in the initial BA process allows for higher spectral efficiency and lower power consumption by efficiently turning off some RF chains \cite{heng_six_2021}. 

This paper proposes location- and orientation-aware methods to reduce the overhead of initial BA for multi-panel devices by recommending a short beam/panel candidate list. We generalize the single-network (SN) structure we proposed in \cite{rezaie_location-_2020} to handle hybrid beamforming at UT. In addition, two multi-network (MN) designs especially tailored to multi-panel devices are presented in this work. MN designs containing fewer trainable parameters can outperform SN design, especially when trained with datasets of limited size. We consider two baselines for evaluations: generalized IFP (GIFP) as a benchmark using context information and hierarchical panel-beam selection (HP-BS) as a context-free approach. We use 3-dimensional (3D) ray tracing modeling of the channel responses using an IEEE standard indoor environment. In our simulations, we use patch antenna elements with radiation pattern following a 3GPP model. The results demonstrate the performance improvement in latency and achievable rate using the proposed deep learning (DL)-based methods over the baselines. This study considers static and mobile blockers like humans in the environment, while the impact of hand blockage on the initial BA process is left for future work. 

\section{System and Channel Model}\label{Sec:SysModel}
\begin{figure}[t]
	\centering
	\subfloat[ Edge design\label{1a}]{%
       \tdplotsetmaincoords{-25}{0}
	\scalebox{0.25}{
	\begin{tikzpicture}[tdplot_main_coords]
    
	\pgfmathsetmacro{\cubex}{7}
	\pgfmathsetmacro{\cubey}{14}
	\pgfmathsetmacro{\cubez}{-1}
	
	\pgfmathsetmacro{\patchl}{3}
	\pgfmathsetmacro{\patchw}{0.8}
	\pgfmathsetmacro{\patchs}{1.7}
	
	
	\begin{scope}[rotate around y=50]
	
	\draw[color=darkcerulean, line width = 3pt] (0,0,0) -- ++(\cubex,0,0) -- ++(0,\cubey,0) -- ++(-\cubex,0,0) -- cycle;
	\draw[color=darkcerulean, line width = 3pt] (0,0,\cubez) -- ++(0,\cubey,0) -- ++(\cubex,0,0);
	\draw[color=darkcerulean, line width = 3pt, dash pattern=on 10pt off 12pt, opacity=0.4] (0,0,\cubez) -- ++(\cubex,0,0) -- ++(0,\cubey,0);
	\draw[color=darkcerulean, line width = 3pt] (0,0,0) -- ++(0,0,\cubez);
	\draw[color=darkcerulean, line width = 3pt] (0,\cubey,0) -- ++(0,0,\cubez);
	\draw[color=darkcerulean, line width = 3pt] (\cubex,\cubey,0) -- ++(0,0,\cubez);
	\draw[color=darkcerulean, line width = 3pt, dash pattern=on 10pt off 10pt, opacity=0.4] (\cubex,0,0) -- ++(0,0,\cubez);

    \draw[color=darkgreen, fill=applegreen, line width = 3pt] (0,\cubey/2 - \patchl/2,-0.1) -- ++(0,\patchl,0) -- ++(0,0,-\patchw) -- ++(0,-\patchl,0) -- cycle;
    \node[scale=3] at (-2, \cubey/2+0.5,\cubez/2) {P$3$};
    
    \draw[color=darkgreen, fill=applegreen, line width = 3pt] (\cubex/2 - \patchl/2,\cubey,-0.1) -- ++(\patchl,0,0) -- ++(0,0,-\patchw) -- ++(-\patchl,0,0) -- cycle;
    \node[scale=3] at (\cubex/2, \cubey+0.7,\cubez) {P$2$};
    
    \draw[color=darkgreen, fill=darkgray, line width = 3pt] (\cubex,\cubey/2 - \patchl/2,-0.1) -- ++(0,\patchl,0) -- ++(0,0,-\patchw) -- ++(0,-\patchl,0) -- cycle;
    \node[scale=3] at (\cubex+2, \cubey/2-0.5,\cubez/2) {P$1$};
    
    \draw[>=latex',  ->, line width  = 2pt] (\cubex/2,\cubey/2,0) -- ++ (2,0,0) node[above left =-6pt and -10pt, scale=2]{$x$};  
    \draw[>=latex',  ->, line width  = 2pt] (\cubex/2,\cubey/2,0) -- ++ (0,2,0) node[above right=-5pt and -15pt, scale=2]{$y$};  
    \draw[>=latex',  ->, line width  = 2pt] (\cubex/2,\cubey/2,0) -- ++ (0,0,2) node[above right=-15pt and -7pt, scale=2]{$z$};  
    
    \draw[>=latex',  ->, line width = 2pt, red] (\cubex/2,\cubey/2,0) -- ++ (-0.87,-0.1,1.8) node{};  
    \draw[dash pattern=on 5pt off 5pt, line width  = 2pt, red] (\cubex/2,\cubey/2,0) -- ++ (-0.87,-0.6,0) --++ (0,0.5,1.8);  
    
    \tdplotdrawarc[color=blue, ->, line width = 2pt, dash pattern=on 5pt off 5pt]{(\cubex/2,\cubey/2,0)}{1}{0}{215}{above left =-35pt and 3pt,color=blue, scale=2}{$\phi$}
    
    \tdplotsetthetaplanecoords{215}
    \tdplotdrawarc[tdplot_rotated_coords, color=blue, ->, line width = 2pt, dash pattern=on 5pt off 5pt]{(-12.9,-4.5,0)}{2}{0}{24}{below = -7pt, scale=2}{$\theta$}
    
    \end{scope}
    
	\end{tikzpicture} }}
    \hspace{4em}%
  \subfloat[Edge-face design\label{1b}]{%
       \tdplotsetmaincoords{-25}{0}
	\scalebox{0.25}{
	\begin{tikzpicture}[tdplot_main_coords]
    
	\pgfmathsetmacro{\cubex}{7}
	\pgfmathsetmacro{\cubey}{14}
	\pgfmathsetmacro{\cubez}{-1}
	
	\pgfmathsetmacro{\patchl}{3}
	\pgfmathsetmacro{\patchw}{0.8}
	\pgfmathsetmacro{\patchs}{1.7}
	
	
	\begin{scope}[rotate around y=50]
	
	\draw[color=darkgreen, fill=darkgray, line width = 3pt] (-2,13.2,-\cubez) -- ++(0,\patchs,0) -- ++(\patchs,0,0) -- ++(0,-\patchs,0) -- cycle;
	\node[scale=3] at (-1, \cubey-1,\cubez) {P$4$};
	
	\draw[color=darkcerulean, line width = 3pt] (0,0,0) -- ++(\cubex,0,0) -- ++(0,\cubey,0) -- ++(-\cubex,0,0) -- cycle;
	\draw[color=darkcerulean, line width = 3pt] (0,0,\cubez) -- ++(0,\cubey,0) -- ++(\cubex,0,0);
	\draw[color=darkcerulean, line width = 3pt, dash pattern=on 10pt off 12pt, opacity=0.4] (0,0,\cubez) -- ++(\cubex,0,0) -- ++(0,\cubey,0);
	\draw[color=darkcerulean, line width = 3pt] (0,0,0) -- ++(0,0,\cubez);
	\draw[color=darkcerulean, line width = 3pt] (0,\cubey,0) -- ++(0,0,\cubez);
	\draw[color=darkcerulean, line width = 3pt] (\cubex,\cubey,0) -- ++(0,0,\cubez);
	\draw[color=darkcerulean, line width = 3pt, dash pattern=on 10pt off 10pt, opacity=0.4] (\cubex,0,0) -- ++(0,0,\cubez);

    \draw[color=darkgreen, fill=applegreen, line width = 3pt] (0,\cubey/2 - \patchl/2,-0.1) -- ++(0,\patchl,0) -- ++(0,0,-\patchw) -- ++(0,-\patchl,0) -- cycle;
    \node[scale=3] at (-2, \cubey/2+0.5,\cubez/2) {P$3$};
    
    \draw[color=darkgreen, fill=applegreen, line width = 3pt] (\cubex/2 - \patchl/2,\cubey,-0.1) -- ++(\patchl,0,0) -- ++(0,0,-\patchw) -- ++(-\patchl,0,0) -- cycle;
    \node[scale=3] at (\cubex/2, \cubey+0.7,\cubez) {P$2$};
    
    \draw[color=darkgreen, fill=darkgray, line width = 3pt] (\cubex,\cubey/2 - \patchl/2,-0.1) -- ++(0,\patchl,0) -- ++(0,0,-\patchw) -- ++(0,-\patchl,0) -- cycle;
    \node[scale=3] at (\cubex+2, \cubey/2-0.5,\cubez/2) {P$1$};
    
    \draw[color=darkgreen, fill=applegreen, line width = 3pt] (\cubex-2.5
    ,0.5,0.4) -- ++(0,\patchs,0) -- ++(\patchs,0,0) -- ++(0,-\patchs,0) -- cycle;
    \node[scale=3] at (\cubex+1,0.5) {P$5$};
    
    \draw[>=latex',  ->, line width  = 2pt] (\cubex/2,\cubey/2,0) -- ++ (2,0,0) node[above left =-6pt and -10pt, scale=2]{$x$};  
    \draw[>=latex',  ->, line width  = 2pt] (\cubex/2,\cubey/2,0) -- ++ (0,2,0) node[above right=-5pt and -15pt, scale=2]{$y$};  
    \draw[>=latex',  ->, line width  = 2pt] (\cubex/2,\cubey/2,0) -- ++ (0,0,2) node[above right=-15pt and -7pt, scale=2]{$z$};  
    
    \draw[>=latex',  ->, line width = 2pt, red] (\cubex/2,\cubey/2,0) -- ++ (-0.87,-0.1,1.8) node{};  
    \draw[dash pattern=on 5pt off 5pt, line width  = 2pt, red] (\cubex/2,\cubey/2,0) -- ++ (-0.87,-0.6,0) --++ (0,0.5,1.8);  
    
    \tdplotdrawarc[color=blue, ->, line width = 2pt, dash pattern=on 5pt off 5pt]{(\cubex/2,\cubey/2,0)}{1}{0}{215}{above left =-35pt and 3pt,color=blue, scale=2}{$\phi$}
    
    \tdplotsetthetaplanecoords{215}
    \tdplotdrawarc[tdplot_rotated_coords, color=blue, ->, line width = 2pt, dash pattern=on 5pt off 5pt]{(-12.9,-4.5,0)}{2}{0}{24}{below = -7pt, scale=2}{$\theta$}
    
    \end{scope}
	\end{tikzpicture} }}
	\caption{Antenna placement designs (a) Edge design with $3$ ULAs on the device's edges (b) Edge-face design which has $2$ extra UPAs on the device's face and back. Also, the UT LCS is depicted.}
	\label{Fig:Designs}
\end{figure}
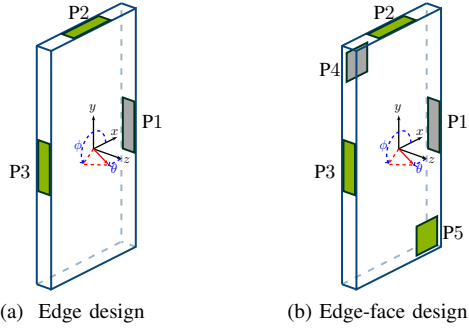

A single-panel fixed AP and a multi-panel mobile UT are located in an indoor 3D scenario. The AP panel is made of a standard uniform planar array (UPA) with size of $\{N_\mathrm{AP_x}, N_\mathrm{AP_y}, N_\mathrm{AP_z}\}$ including $N_\mathrm{AP} = N_\mathrm{AP_x} N_\mathrm{AP_y} N_\mathrm{AP_z}$ elements. The UT has $N_{P}$ panels where each panel is made of a standard uniform linear array (ULA) or UPA in each module. The $p$th panel of the UT consists of a $\{N_\mathrm{UT_x}^{(p)}, N_\mathrm{UT_y}^{(p)}, N_\mathrm{UT_z}^{(p)}\}$ antenna array with $N_\mathrm{UT}^{(p)} = N_\mathrm{UT_x}^{(p)} N_\mathrm{UT_y}^{(p)} N_\mathrm{UT_z}^{(p)}$ elements. The total number of UT antenna elements is thus $N_\mathrm{UT} = \sum_{p=1}^{N_{P}} N_\mathrm{UT}^{(p)}$. Fig. \ref{Fig:Designs} shows the edge and edge-face designs of antenna placement at UT, inspired from \cite{raghavan_antenna_2019}. The edge design contains $3$ ULA panels, while the edge-face design includes $2$ additional UPA modules on the device's face and back.

The position and orientation of the AP and UT are defined in a global coordinate system (GCS) \cite{ali_orientation-assisted_2021}. Both the AP and UT have their own local coordinate systems (LCS) such that the AP's UPA and the UT's screen are oriented parallel to the $yz$ and $xy$ plane of their respective LCSs. The AP is placed at $\boldsymbol{p}_\mathrm{AP} = (x_\mathrm{AP}, y_\mathrm{AP}, z_\mathrm{AP}) \in \mathbb{R}^3$ with their LCSs rotated by angles $\boldsymbol{\psi}_\mathrm{AP} = (\alpha_\mathrm{AP}, \beta_\mathrm{AP}, \gamma_\mathrm{AP})$ around $z$, $y$, and $x$ axes of the GCS. The UT takes a random position in the environment at position  $\boldsymbol{p}_\mathrm{UT} = (x_\mathrm{UT}, y_\mathrm{UT}, z_\mathrm{UT}) \in \mathbb{R}^3$ with rotation vector $\boldsymbol{\psi}_\mathrm{UT} = (\alpha_\mathrm{UT}, \beta_\mathrm{UT}, \gamma_\mathrm{UT})$ defined analogously to $\boldsymbol{\psi}_\mathrm{AP}$ \cite{rezaie_deep_2021-1}. Two modes of device orientation are considered in this study: portrait and landscape. In portrait mode, $\beta_\mathrm{UT} = 0$ and $\alpha_\mathrm{UT}$,  $\gamma_\mathrm{UT}$ are uniformly random in the ranges $\alpha_\mathrm{UT} \in [-\pi, \pi)$ and $\gamma_\mathrm{UT} \in [0, \pi/2]$. In the landscape mode, $\gamma_\mathrm{UT} = 0$ and $\alpha_\mathrm{UT}$,  $\beta_\mathrm{UT}$ are drawn uniformly in the ranges $\alpha_\mathrm{UT} \in [-\pi, \pi)$ and $\beta_\mathrm{UT} \in [-\pi/2, 0]$.

\subsection{Channel and Signal Model}
We use Altair WinProp\textsuperscript{TM} as ray-tracing software \cite{noauthor_altair_nodate} to generate channel responses between AP and UT panels at each UT location and orientation. The channel to each UT pannel is built using outputs of the tool such as angle of departure (AoD), angle of arrival (AoA), and gains of all paths between the AP and the panel. The channel matrix $\boldsymbol{H}^{(p)} \in \mathbb{C}^{N_\mathrm{UT}^{(p)} \times N_\mathrm{AP}}$ between the AP and the $p$th UT panel is modeled as
\begin{equation}\label{Eq:Channel}
    \boldsymbol{H}^{(p)} = \sum_{l=0}^{L^{(p)}} \sqrt{\rho_l^{(p)}} \hspace{2pt} e^{j\vartheta_l^{(p)}} \hspace{2pt} \boldsymbol{a}_\mathrm{UT}^{(p)}(\phi^{(p)}_{l}, \theta^{(p)}_{l}) \hspace{2pt} \boldsymbol{a}_\mathrm{AP}^H(\psi^{(p)}_{l},\omega^{(p)}_{l})
\end{equation}
where $L^{(p)}$, $\rho_l^{(p)}$, and $\vartheta_l^{(p)}$ denote the number of multipath components seen at the $p$th panel, the received power, and the phase of the $l$th path, respectively. $\boldsymbol{a}_\mathrm{UT}^{(p)}$ and $\boldsymbol{a}_\mathrm{AP}$ show the antenna array response of the $p$th UT panel and the AP. Also, $\phi^{(p)}_{l}$ and $\theta^{(p)}_{l}$ are the azimuth and elevation AoAs of the $l$th path in the UT LCS, respectively. Likewise, $\psi^{(p)}_{l}$ and $\omega^{(p)}_{l}$ denote the azimuth and elevation AoDs of the $l$th path in the AP LCS. The array response of a $\{N_\mathrm{x}, N_\mathrm{y}, N_\mathrm{z}\}$ antenna array with $N_\mathrm{a} = N_\mathrm{x} N_\mathrm{y} N_\mathrm{z}$ elements may be written as
\begin{equation}
    \boldsymbol{a}(\phi, \theta) = \frac{1}{\sqrt{N_\mathrm{a}}} g_\mathrm{a}(\phi, \theta) \boldsymbol{a}_\mathrm{z}(\theta) \otimes \boldsymbol{a}_\mathrm{y}(\phi, \theta) \otimes \boldsymbol{a}_\mathrm{x}(\phi, \theta)
\end{equation}
where $g_\mathrm{a}(\phi, \theta)$ is the antenna gain at azimuth and elevation angles $\phi$ and $\theta$, and 
$\boldsymbol{a}_\mathrm{x} \in \mathbb{C}^{N_\mathrm{x} \times 1}$, $\boldsymbol{a}_\mathrm{y} \in \mathbb{C}^{N_\mathrm{y} \times 1}$, and $\boldsymbol{a}_\mathrm{z} \in \mathbb{C}^{N_\mathrm{z} \times 1}$ are given by
\begin{equation}\label{eq:a_x}
     \boldsymbol{a}_\mathrm{x}(\phi, \theta) =  [1 , e^{j \pi sin(\theta) \hspace{2pt} cos(\phi) }, \dots 
     , e^{ j \pi (N_\mathrm{x} - 1) \hspace{2pt} sin(\theta) \hspace{2pt} cos(\phi) }]^T,
\end{equation}
\begin{equation}\label{eq:a_y}
     \boldsymbol{a}_\mathrm{y}(\phi, \theta) =  [1 , e^{j \pi sin(\theta) \hspace{2pt} sin(\phi) }, \dots 
     , e^{ j \pi (N_\mathrm{y} - 1) \hspace{2pt} sin(\theta) \hspace{2pt} sin(\phi) }]^T,
\end{equation}
\begin{equation}\label{eq:a_z}
     \boldsymbol{a}_\mathrm{z}(\theta) =  [1 , e^{j \pi cos(\theta)}, \dots 
     , e^{ j \pi (N_\mathrm{z} - 1) \hspace{2pt} cos(\theta) }]^T.
\end{equation}
The received signal at the $p$th UT panel may be written as
\begin{equation}\label{Y_Rec_Sig}
    \boldsymbol{y}^{(p)}= \sqrt{P_\mathrm{AP}} {\boldsymbol{v}^{(p)}}^H \boldsymbol{H}^{(p)} \boldsymbol{u} s + {\boldsymbol{v}^{(p)}}^H \boldsymbol{n}^{(p)}
\end{equation}
where $P_\mathrm{AP}$, $s$, and $\boldsymbol{u}$ respectively denote the transmission power, the unit power transmitted symbol, and the beamforming vector used at the AP. $\boldsymbol{v}^{(p)}$, $\boldsymbol{n}^{(p)}$ are, respectively, the combiner and a zero-mean complex Gaussian noise vector with variance $\sigma_n^2$ at the $p$th panel of the UT. 
\subsection{Single-Panel Codebook}
This study considers analog phased antenna arrays at the AP and each UT panel. $N_{RF}$ denotes the number of RF chains at the UT. When $N_{RF}>1$, simultaneous sensing with different panels can be done.  For simplicity, however, we do not consider multi-panel beamforming in this study. In addition, discrete Fourier transform (DFT)-based codebooks using analog phase shifters are used \cite{rezaie_location-_2020}. We consider the set $\boldsymbol{\mathcal{U}} = \{\boldsymbol{u}_1, \dots, \boldsymbol{u}_{N_\mathrm{AP}}\}$ including all the accessible beamforming vectors at the AP. For the $p$th UT panel, we define $\boldsymbol{\mathcal{V}}^{p} = \{\boldsymbol{v}^{p}_{1}, \dots, \boldsymbol{v}^{p}_{{N^{(p)}_\mathrm{UT}}}\}$ as the panel codebook. The set $\boldsymbol{\mathcal{V}} = \{\boldsymbol{v}_1, \dots, \boldsymbol{v}_{N_\mathrm{UT}}\}$ includes all combiners at the UT, as the union of all panel codebooks. The received signal strength (RSS) using the beamforming vector $\boldsymbol{u}_i$ and combiner $\boldsymbol{v}_j$ at the AP and UT can be written as
\begin{equation}\label{Eq:RSS}
    R_{i, j} = \Big | \sqrt{P_\mathrm{AP}} \boldsymbol{v}_j^H \boldsymbol{H}^{(p)} \boldsymbol{u}_i s + \boldsymbol{v}_j^H \boldsymbol{n}  \Big | ^2
\end{equation}
where $p$ denotes the panel corresponding  to combiner $\boldsymbol{v}_j$.

\section{Deep Learning based Beam Selection}
Let the set $\mathcal{B}$ include the indices of all possible combinations of beamforming vectors and combiners in the AP and UT. Beam ranking, as a beam selection approach for a UT with a known position and orientation, can be seen as an optimization problem in finding the subset $\mathcal{S}$ from $\mathcal{B}$ which minimizes the misalignment probability, i.e., 
\begin{equation} \label{eq:Mis}
\begin{aligned}
    \min_{\mathcal{S}} \quad & \mathbb{P} \left[\max\limits_{(t, w) \in \mathcal{S}} R_{t, w} < \max\limits_{(i, j) \in \mathcal{B}} R_{i,j}\right], \\
    \textrm{s.t.} \quad & | \mathcal{S} | = C
\end{aligned}
\end{equation}
where $C$ is a pre-defined constant. The optimal AP/UT beam pair for transmission is
\begin{equation} \label{eq:i_j_max}
    i^\star, j^\star = \underset{(i, j) \in\mathcal{B}}{\mathrm{arg\hspace{2pt}max}} \hspace{2pt} R_{i, j}.
\end{equation}
In addition, we define $p^\star$ as the panel corresponding to the optimal UT beam $j^\star$. $P_{i,j} = \mathbb{P} \left[ (i,j) = (i^\star, j^\star) \right]$\footnote{$P_{i,j}$ is a conditional probability given the UT location and orientation. To alleviate the notation, we drop this conditionality from the equations.} denotes the probability of beam pair $(i,j)$ being optimal, i.e., being the beam pair in the codebook yielding the highest RSS. 

The AP and UT sense the environment with the beam pairs included in $\mathcal{S}$, and the UT reports the beam pair $(\hat{i}, \hat{j})$ with highest RSS as a result of the beam selection procedure. For known UT location and orientation, the optimal $\mathcal{S}$ includes beam pairs with the highest probabilities of optimality \cite{va_inverse_2018}. We present next our proposed deep learning methods, which estimate optimality probabilities of all beam pairs. 

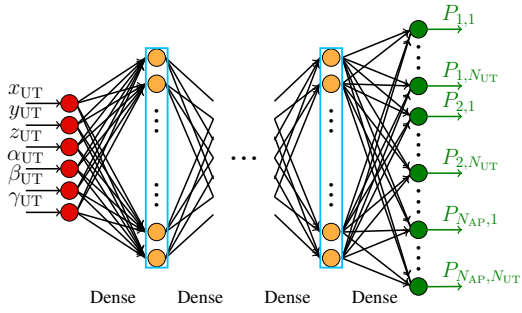
\begin{figure}[!t]
	\centering
	\definecolor{mycolor1}{RGB}{224,9,2} 
	\definecolor{mycolor2}{RGB}{255,174,66} 
	\definecolor{mycolor3}{RGB}{0,120,210} 
	\definecolor{mycolor4}{RGB}{0, 128, 0} 
	\definecolor{mycolor5}{RGB}{0,191,255} 
	\definecolor{mycolor6}{RGB}{0,0,139} 
	\definecolor{mycolor7}{RGB}{255,0,255} 
	\definecolor{mycolor8}{RGB}{238,130,238} 
	\definecolor{mycolor9}{RGB}{128,0,128} 
	\scalebox{.58}{
		\begin{tikzpicture}[cross/.style={path picture={ 
				\draw[red, line width=3pt]
				(path picture bounding box.south east) -- (path picture bounding box.north west) (path picture bounding box.south west) -- (path picture bounding box.north east);
		}}]
		\foreach \i in {-2,...,3} {
			\foreach \h in {0,...,-1} {
				\draw[black, ->-=0.95, line width=1pt](0.15 , 0.5*\i-0.5*0.5) -- (1.75 ,0.6 * \h - 1.7);
				\draw[black, ->-=0.95, line width=1pt](0.15 , 0.5*\i-0.5*0.5) -- (1.75 ,-0.6 * \h + 1.7);
			}
		}

		\foreach \i in {0,...,-1} {
			\foreach \h in {-1,...,1} {
				\draw[black, line width=1pt](2.2 ,0.6 * \i - 1.7) -- (3.8-0.5 , 1.3 * \h  - 0.5/1.6*1.3*\h +0.5/1.6*1.3*\i);
				\draw[black, line width=1pt](2.2 ,-0.6 * \i + 1.7) -- (3.8-0.5 , 1.3 * \h  - 0.5/1.6*1.3*\h -0.5/1.6*1.3*\i);
			}
		}

		\foreach \i in {-1,...,1} {
			\foreach \h in {0,...,-1} {
				\draw[black, ->-=0.95, line width=1pt](4.2 +0.5,1.3 * \i + 0.5*1.3*\h/1.6 - 0.5*1.3*\i/1.6) -- (5.75 ,0.6 * \h - 1.7);
				\draw[black, ->-=0.95, line width=1pt](4.2 + 0.5,1.3 * \i - 0.5*1.3*\h/1.6 - 0.5*1.3*\i/1.6) -- (5.75 ,-0.6 * \h + 1.7);
			}
		}

		\foreach \i in {0,...,-1} {
			\foreach \h in {-2,...,1} {
                \draw[black, ->-=0.95, line width=1pt](6.25 ,0.6 * \i - 1.7) -- (7.85 ,1.3 * \h- 0.35);
                \draw[black, ->-=0.95, line width=1pt](6.25 ,-0.6 * \i + 1.7) -- (7.85 ,1.3 * \h- 0.35);
			}
		}
		
		\foreach \i in {0,...,-1} {
			\foreach \h in {1,...,2} {
				\draw[black, ->-=0.95, line width=1pt](6.25 ,0.6 * \i - 1.7) -- (7.85 ,1.3 * \h+ 0.35);
				\draw[black, ->-=0.95, line width=1pt](6.25 ,-0.6 * \i + 1.7) -- (7.85 ,1.3 * \h+ 0.35);
			}
		}
		
		\path (3.8,0) -- (4.2,0) node [black, font=\Huge, midway, sloped] {\textbf{$...$}};
		
		\foreach \i in {-2,...,3} {
    		\draw[fill=mycolor1] (0,0.5*\i-0.5*0.5) circle (0.2);  
    	}
		
		\draw[fill=mycolor2] (2,-2.3) circle (0.2);  
		\draw[fill=mycolor2] (2,-1.7) circle (0.2);  
		\path (2,-1.7) -- (2,0) node [black, font=\Huge, midway, sloped] {$...$};
		\path (2,0) -- (2,1.7) node [black, font=\Huge, midway, sloped] {$...$};
		\draw[fill=mycolor2] (2,1.7) circle (0.2);  
		\draw[fill=mycolor2] (2,2.3) circle (0.2);  
		
		\coordinate (dm1) at (2,-2.4);
        \coordinate (dm2) at (2,2.4);
        \node[rectangle, draw, mycolor5, very thick, minimum width=0.5cm] [fit = (dm1) (dm2)] (bx4) {};
		\node[align=center,font=\large,rotate=90] at (bx4.center) {};
		
		\draw[fill=mycolor2] (6,-2.3) circle (0.2);  
		\draw[fill=mycolor2] (6,-1.7) circle (0.2);  
		\path (6,-1.7) -- (6,0) node [black, font=\Huge, midway, sloped] {$...$};
		\path (6,0) -- (6,1.7) node [black, font=\Huge, midway, sloped] {$...$};
		\draw[fill=mycolor2] (6,1.7) circle (0.2);  
		\draw[fill=mycolor2] (6,2.3) circle (0.2);  
		
		\coordinate (dm1) at (6,-2.4);
        \coordinate (dm2) at (6,2.4);
        \node[rectangle, draw, mycolor5, very thick, minimum width=0.5cm] [fit = (dm1) (dm2)] (bx4) {};
		\node[align=center,font=\large,rotate=90] at (bx4.center) {};
		
		\draw[fill=mycolor4] (8,-2.95) circle (0.2);  
		\path (8,-2.95) -- (8,-1.65) node [black, font=\Huge, midway, sloped] {$...$};
		\draw[fill=mycolor4] (8,-1.65) circle (0.2);  
		\path (8,-1.65) -- (8,-0.35) node [black, font=\Huge, midway, sloped] {$...$};
		\draw[fill=mycolor4] (8,-0.35) circle (0.2);  
		\path (8,-0.35) -- (8,0.95) node [black, font=\Huge, midway, sloped] {$...$};
		\draw[fill=mycolor4] (8,0.95) circle (0.2);  
		\draw[fill=mycolor4] (8,1.65) circle (0.2);  
		\path (8,1.65) -- (8,2.95) node [black, font=\Huge, midway, sloped] {\textbf{$...$}};
		\draw[fill=mycolor4] (8,2.95) circle (0.2);  
		
		\draw[->, line width=1pt] (-1,-0.5*2-0.5*0.5) node [font=\Large, above] {$\gamma_\mathrm{UT}$} -- (-0.2,-0.5*2-0.5*0.5);
		\draw[->, line width=1pt] (-1,-0.5*1-0.5*0.5) node [font=\Large, above] {$\beta_\mathrm{UT}$} -- (-0.2,-0.5*1-0.5*0.5);
		\draw[->, line width=1pt] (-1,-0.5*0-0.5*0.5) node [font=\Large, above] {$\alpha_\mathrm{UT}$} -- (-0.2,-0.5*0-0.5*0.5);
		\draw[->, line width=1pt] (-1,0.5*1-0.5*0.5) node [font=\Large, above] {$z_\mathrm{UT}$} -- (-0.2,0.5*1-0.5*0.5);
		\draw[->, line width=1pt] (-1,0.5*2-0.5*0.5) node [font=\Large, above] {$y_\mathrm{UT}$} -- (-0.2,0.5*2-0.5*0.5);
		\draw[->, line width=1pt] (-1,0.5*3-0.5*0.5) node [font=\Large, above] {$x_\mathrm{UT}$} -- (-0.2,0.5*3-0.5*0.5);

		\draw[->, line width=1pt, mycolor4] (8.15,-2.955) -- (9,-2.95) node [font=\Large, above right = -2pt and -17pt] {$P_{N_\mathrm{AP}, N_\mathrm{UT}}$};
		\draw[->, line width=1pt, mycolor4] (8.15,-1.65) -- (9,-1.65) node [font=\Large, above right = -2pt and -17pt] {$P_{N_\mathrm{AP}, 1}$};
		\draw[->, line width=1pt, mycolor4] (8.15,-0.35) -- (9,-0.35) node [font=\Large, above right = -2pt and -17pt] {$P_{2, N_\mathrm{UT}}$};
		\draw[->, line width=1pt, mycolor4] (8.15,0.95) -- (9,0.95) node [font=\Large, above right = -2pt and -17pt] {$P_{2, 1}$};
		\draw[->, line width=1pt, mycolor4] (8.15,1.65) -- (9,1.65) node [font=\Large, above right = -2pt and -17pt] {$P_{1, N_\mathrm{UT}}$};
		\draw[->, line width=1pt, mycolor4] (8.15,2.95) -- (9,2.95) node [font=\Large, above right = -2pt and -17pt] {$P_{1, 1}$};
		
		\node[align=left] at (1,-3.2) {\large Dense};
		\node[align=left] at (3,-3.2) {\large Dense};
		\node[align=left] at (5,-3.2) {\large Dense};
		\node[align=left] at (7,-3.2) {\large Dense};
		\end{tikzpicture}}
	\caption{The single-network structure directly estimates the optimality probability of each beam pair.} 
	\label{Fig:NetStruc1}
\end{figure}
\subsection{Single-Network Design}\label{SN}
In this structure, the UT location and orientation are fed to a network with $N_h$ hidden layers with $n_h$ neurons each. There are $N_\mathrm{UT} N_\mathrm{AP}$ neurons at the output layer, each neuron yielding an estimate of $P_{i,j}$ for the $(i, j)$th beam pair. The network includes $7 n_h + (N_h-1) (n_h + 1) n_h + (n_h+1) N_\mathrm{UT} N_\mathrm{AP}$ trainable parameters \cite{rezaie_deep_2021-1}. After sorting the network's outputs, a beam candidate list $\mathcal{S}$ including the first $N_b$ indices of beam pairs is made. The AP communicates the combiners selected in $\mathcal{S}$, and then the transceivers  sense the environment using $N_b$ time slots during the BA phase \cite{rezaie_location-_2020}. Originally, this method was developed for analog beamforming at UT. In this paper, we generalize this method to propose a beam candidate list accounting for hybrid beamforming at UT. This is achieved by updating the beam candidate list to sense multiple beam pairs at each time slot. For each member of set $\mathcal{S}$, $N_{RF}-1$ beam pairs from different panels are added to the final candidate list. These beam pairs provide higher probabilities of being optimal with the same AP beam. Thus, $\mathcal{S}^{new}$ includes $N_{RF} N_b$ beam pairs that can be sensed in $N_b$ time slots.

\subsection{Multi-Network Panel Selection Design}
\begin{figure}[t]
	\centering
	\subfloat[ $\mathit{NET}^P_{\mathrm{I}}$\label{I}]{%
    \centering
	\definecolor{mycolor1}{RGB}{224,9,2} 
	\definecolor{mycolor2}{RGB}{255,174,66} 
	\definecolor{mycolor3}{RGB}{0,120,210} 
	\definecolor{mycolor4}{RGB}{0, 128, 0} 
	\definecolor{mycolor5}{RGB}{0,191,255} 
	\definecolor{mycolor6}{RGB}{0,0,139} 
	\definecolor{mycolor7}{RGB}{255,0,255} 
	\definecolor{mycolor8}{RGB}{238,130,238} 
	\definecolor{mycolor9}{RGB}{128,0,128} 
	\scalebox{.58}{
		\begin{tikzpicture}[cross/.style={path picture={ 
				\draw[red, line width=3pt]
				(path picture bounding box.south east) -- (path picture bounding box.north west) (path picture bounding box.south west) -- (path picture bounding box.north east);
		}}]
		\foreach \i in {-1,...,1} {
			\foreach \h in {0,...,-1} {
				\draw[black, ->-=0.95, line width=1pt](0.15 , 0.5*\i) -- (1.75 ,0.6 * \h - 1.7);
				\draw[black, ->-=0.95, line width=1pt](0.15 , 0.5*\i) -- (1.75 ,-0.6 * \h + 1.7);
			}
		}
		
		\foreach \i in {0,...,-1} {
			\foreach \h in {-1,...,1} {
				\draw[black, line width=1pt](2.2 ,0.6 * \i - 1.7) -- (3.8-0.5 , 1.3 * \h  - 0.5/1.6*1.3*\h +0.5/1.6*1.3*\i);
				\draw[black, line width=1pt](2.2 ,-0.6 * \i + 1.7) -- (3.8-0.5 , 1.3 * \h  - 0.5/1.6*1.3*\h -0.5/1.6*1.3*\i);
			}
		}

		\foreach \i in {-1,...,1} {
			\foreach \h in {0,...,-1} {
				\draw[black, ->-=0.95, line width=1pt](4.2 +0.5,1.3 * \i + 0.5*1.3*\h/1.6 - 0.5*1.3*\i/1.6) -- (5.75 ,0.6 * \h - 1.7);
				\draw[black, ->-=0.95, line width=1pt](4.2 + 0.5,1.3 * \i - 0.5*1.3*\h/1.6 - 0.5*1.3*\i/1.6) -- (5.75 ,-0.6 * \h + 1.7);
			}
		}

		\foreach \i in {0,...,-1} {
			\foreach \h in {-1,...,-1} {
                \draw[black, ->-=0.95, line width=1pt](6.25 ,0.6 * \i - 1.7) -- (7.85 ,1.3 * \h- 0.35);
                \draw[black, ->-=0.95, line width=1pt](6.25 ,-0.6 * \i + 1.7) -- (7.85 ,1.3 * \h- 0.35);
			}
		}
		
		\foreach \i in {0,...,-1} {
			\foreach \h in {0,...,1} {
				\draw[black, ->-=0.95, line width=1pt](6.25 ,0.6 * \i - 1.7) -- (7.85 ,0.7 * \h+ 0.95);
				\draw[black, ->-=0.95, line width=1pt](6.25 ,-0.6 * \i + 1.7) -- (7.85 ,0.7 * \h+ 0.95);
			}
		}
		
		\path (3.8,0) -- (4.2,0) node [black, font=\Huge, midway, sloped] {\textbf{$...$}};
		
		\foreach \i in {-1,...,1} {
    		\draw[fill=mycolor1] (0,0.5*\i) circle (0.2);  
    	}
		
		\draw[fill=mycolor2] (2,-2.3) circle (0.2);  
		\draw[fill=mycolor2] (2,-1.7) circle (0.2);  
		\path (2,-1.7) -- (2,0) node [black, font=\Huge, midway, sloped] {$...$};
		\path (2,0) -- (2,1.7) node [black, font=\Huge, midway, sloped] {$...$};
		\draw[fill=mycolor2] (2,1.7) circle (0.2);  
		\draw[fill=mycolor2] (2,2.3) circle (0.2);  
		
		\coordinate (dm1) at (2,-2.4);
        \coordinate (dm2) at (2,2.4);
        \node[rectangle, draw, mycolor5, very thick, minimum width=0.5cm] [fit = (dm1) (dm2)] (bx4) {};
		\node[align=center,font=\large,rotate=90] at (bx4.center) {};
		
		\draw[fill=mycolor2] (6,-2.3) circle (0.2);  
		\draw[fill=mycolor2] (6,-1.7) circle (0.2);  
		\path (6,-1.7) -- (6,0) node [black, font=\Huge, midway, sloped] {$...$};
		\path (6,0) -- (6,1.7) node [black, font=\Huge, midway, sloped] {$...$};
		\draw[fill=mycolor2] (6,1.7) circle (0.2);  
		\draw[fill=mycolor2] (6,2.3) circle (0.2);  
		
		\coordinate (dm1) at (6,-2.4);
        \coordinate (dm2) at (6,2.4);
        \node[rectangle, draw, mycolor5, very thick, minimum width=0.5cm] [fit = (dm1) (dm2)] (bx4) {};
		\node[align=center,font=\large,rotate=90] at (bx4.center) {};
		
		\draw[fill=mycolor4] (8,-1.65) circle (0.2);  
		\path (8,-1.65) -- (8,0.95) node [black, font=\Huge, midway, sloped] {$...$};
		\draw[fill=mycolor4] (8,0.95) circle (0.2);  
		\draw[fill=mycolor4] (8,1.65) circle (0.2);  
		
		\draw[->, line width=1pt] (-1,-0.5*1) node [font=\Large, above] {$z_\mathrm{UT}$} -- (-0.2,-0.5*1);
		\draw[->, line width=1pt] (-1,0) node [font=\Large, above] {$y_\mathrm{UT}$} -- (-0.2,0);
		\draw[->, line width=1pt] (-1,0.5*1) node [font=\Large, above] {$x_\mathrm{UT}$} -- (-0.2,0.5*1);

		\draw[->, line width=1pt, mycolor4] (8.15,-1.65) -- (9,-1.65) node [font=\Large, above right = -2pt and -17pt] {$P_{N_\mathrm{AP}}$};
		\draw[->, line width=1pt, mycolor4] (8.15,0.95) -- (9,0.95) node [font=\Large, above right = -2pt and -17pt] {$P_{2}$};
		\draw[->, line width=1pt, mycolor4] (8.15,1.65) -- (9,1.65) node [font=\Large, above right = -2pt and -17pt] {$P_{1}$};
		
		\node[align=left] at (1,-3.2) {\large Dense};
		\node[align=left] at (3,-3.2) {\large Dense};
		\node[align=left] at (5,-3.2) {\large Dense};
		\node[align=left] at (7,-3.2) {\large Dense};
		\end{tikzpicture}}}
		\\[0.05cm]
	\hspace{-1cm}
  \subfloat[$\mathit{NET}^P_{\mathrm{II}}$\label{II}]{%
    \centering
	\definecolor{mycolor1}{RGB}{224,9,2} 
	\definecolor{mycolor2}{RGB}{255,174,66} 
	\definecolor{mycolor3}{RGB}{0,120,210} 
	\definecolor{mycolor4}{RGB}{0, 128, 0} 
	\definecolor{mycolor5}{RGB}{0,191,255} 
	\definecolor{mycolor6}{RGB}{0,0,139} 
	\definecolor{mycolor7}{RGB}{255,0,255} 
	\definecolor{mycolor8}{RGB}{238,130,238} 
	\definecolor{mycolor9}{RGB}{128,0,128} 
	\scalebox{.58}{
		\begin{tikzpicture}[cross/.style={path picture={ 
				\draw[red, line width=3pt]
				(path picture bounding box.south east) -- (path picture bounding box.north west) (path picture bounding box.south west) -- (path picture bounding box.north east);
		}}]
		\foreach \i in {-2,...,3} {
			\foreach \h in {0,...,-1} {
				\draw[black, ->-=0.95, line width=1pt](0.15 , 0.5*\i+0.7) -- (1.75 ,-0.6 * \h + 1.7);
			}
		}
		\foreach \i in {-2,...,3} {
				\draw[black, ->-=0.95, line width=1pt](0.15 , 0.5*\i+0.7) -- (1.75 , 0.4);
		}
		\draw[black, ->-=0.95, line width=1pt](0.15, -0.5*4+0.7) -- (1.75, -0.5*4+0.7);

		\foreach \i in {0,...,-1} {
			\foreach \h in {-1,...,1} {
				\draw[black, line width=1pt](2.2+2 ,0.6 * \i - 1.55) -- (3.8-0.5+2 , 1.3 * \h  - 0.5/1.6*1.3*\h +0.5/1.6*1.3*\i);
				\draw[black, line width=1pt](2.2+2 ,-0.6 * \i + 1.55) -- (3.8-0.5+2 , 1.3 * \h  - 0.5/1.6*1.3*\h -0.5/1.6*1.3*\i);
			}
		}
		
		\foreach \h in {-1,...,1} {
			\draw[black, line width=1pt](2.2+2 ,0.3) -- (3.8-0.5+2 , 1.3 * \h  - 0.5/1.6*1.3*\h +0.5/1.6*1.3*0);
			\draw[black, line width=1pt](2.2+2 ,-0.3) -- (3.8-0.5+2 , 1.3 * \h  - 0.5/1.6*1.3*\h -0.5/1.6*1.3*0);
		}

		\foreach \i in {-1,...,1} {
			\foreach \h in {0,...,-1} {
				\draw[black, ->-=0.95, line width=1pt](4.2+2 +0.5,1.3 * \i + 0.5*1.3*\h/1.6 - 0.5*1.3*\i/1.6) -- (5.75+2 ,0.6 * \h - 1.55);
				\draw[black, ->-=0.95, line width=1pt](4.2 + 0.5+2,1.3 * \i - 0.5*1.3*\h/1.6 - 0.5*1.3*\i/1.6) -- (5.75+2 ,-0.6 * \h + 1.55);
			}
		}

		\foreach \i in {0,...,-1} {
			\foreach \h in {-1,...,-1} {
                \draw[black, ->-=0.95, line width=1pt](6.25+2 ,0.6 * \i - 1.55) -- (7.85+2 ,1 * \h);
                \draw[black, ->-=0.95, line width=1pt](6.25+2 ,-0.6 * \i + 1.55) -- (7.85+2 ,1 * \h);
			}
		}
		
		\foreach \i in {0,...,-1} {
			\foreach \h in {0,...,1} {
				\draw[black, ->-=0.95, line width=1pt](6.25+2 ,0.6 * \i - 1.55) -- (7.85+2 ,0.7 * \h+ 0.3);
				\draw[black, ->-=0.95, line width=1pt](6.25+2 ,-0.6 * \i + 1.55) -- (7.85+2 ,0.7 * \h+ 0.3);
			}
		}
		
		\path (3.8+2,0) -- (4.2+2,0) node [black, font=\Huge, midway, sloped] {\textbf{$...$}};
		
		\foreach \i in {-2,...,3} {
    		\draw[fill=mycolor1] (0,0.5*\i+0.7) circle (0.2);  
    	}
    	\draw[fill=mycolor1] (0,-0.5*4+0.7) circle (0.2);  
		
		\draw[fill=mycolor7] (2,-2.3) circle (0.2);  
		\draw[fill=mycolor7] (2,-1.7) circle (0.2);  
		\path (2,-1.7) -- (2,-0.4) node [black, font=\Huge, midway, sloped] {$...$};
		\draw[fill=mycolor7] (2,-0.4) circle (0.2);  
		\draw[fill=mycolor2] (2,0.4) circle (0.2);  
		\path (2,0.4) -- (2,1.7) node [black, font=\Huge, midway, sloped] {$...$};
		\draw[fill=mycolor2] (2,1.7) circle (0.2);  
		\draw[fill=mycolor2] (2,2.3) circle (0.2);  
		
		\coordinate (dm1) at (2,-.25);
        \coordinate (dm2) at (2,-2.4);
        \node[rectangle, draw, mycolor3, very thick, minimum width=0.5cm] [fit = (dm1) (dm2)] (bx4) {};
		\node[align=center,font=\large,rotate=90] at (bx4.center) {};
		
		\coordinate (dm1) at (2,0.25);
        \coordinate (dm2) at (2,2.4);
        \node[rectangle, draw, mycolor5, very thick, minimum width=0.5cm] [fit = (dm1) (dm2)] (bx4) {};
		\node[align=center,font=\large,rotate=90] at (bx4.center) {};
		
		\draw[fill=mycolor7] (2+2,-2.15) circle (0.2);  
		\draw[fill=mycolor7] (2+2,-1.55) circle (0.2);  
		\path (2+2,-1.55) -- (2+2,-0.25) node [black, font=\Huge, midway, sloped] {$...$};
		\draw[fill=mycolor7] (2+2,-0.25) circle (0.2);  
		\draw[fill=mycolor2] (2+2,0.25) circle (0.2);  
		\path (2+2,0.25) -- (2+2,1.55) node [black, font=\Huge, midway, sloped] {$...$};
		\draw[fill=mycolor2] (2+2,1.55) circle (0.2);  
		\draw[fill=mycolor2] (2+2,2.15) circle (0.2);  
		
		\coordinate (dm1) at (2+2,-0.1);
        \coordinate (dm2) at (2+2,-2.25);
        \node[rectangle, draw, mycolor3, very thick, minimum width=0.5cm] [fit = (dm1) (dm2)] (bx4) {};
		\node[align=center,font=\large,rotate=90] at (bx4.center) {};
		
		\coordinate (dm1) at (2+2,0.1);
        \coordinate (dm2) at (2+2,2.25);
        \node[rectangle, draw, mycolor5, very thick, minimum width=0.5cm] [fit = (dm1) (dm2)] (bx4) {};
		\node[align=center,font=\large,rotate=90] at (bx4.center) {};
		
		\coordinate (A) at (2.25,1.2);
		\coordinate (B) at (2.25,-1.2);
        \coordinate (C) at (3.75,0);
		\draw[line width=0.5mm, dotted] (A) to[out=0,in=180] (C);
		\draw[line width=0.5mm, dotted] (B) to[out=0,in=180] (C);
		
		\draw[fill=mycolor2] (6+2,-2.15) circle (0.2);  
		\draw[fill=mycolor2] (6+2,-1.55) circle (0.2);  
		\path (6+2,-1.55) -- (6+2,0) node [black, font=\Huge, midway, sloped] {$...$};
		\path (6+2,0) -- (6+2,1.55) node [black, font=\Huge, midway, sloped] {$...$};
		\draw[fill=mycolor2] (6+2,1.55) circle (0.2);  
		\draw[fill=mycolor2] (6+2,2.15) circle (0.2);  
		
		\coordinate (dm1) at (6+2,-2.25);
        \coordinate (dm2) at (6+2,2.25);
        \node[rectangle, draw, mycolor5, very thick, minimum width=0.5cm] [fit = (dm1) (dm2)] (bx4) {};
		\node[align=center,font=\large,rotate=90] at (bx4.center) {};
		
		\draw[fill=mycolor4] (8+2,-1) circle (0.2);  
		\path (8+2,-1.65) -- (8+2,0.95) node [black, font=\Huge, midway, sloped] {$...$};
		\draw[fill=mycolor4] (8+2,0.3) circle (0.2);  
		\draw[fill=mycolor4] (8+2,1) circle (0.2);  

		\draw[->, line width=1pt] (-1,-0.5*4+0.7) node [font=\Large, above] {AP beam} -- (-0.2,-0.5*4+0.7);
		\draw[->, line width=1pt] (-1,-0.5*4+0.7) node [font=\Large, below] {index $i$} -- (-0.2,-0.5*4+0.7);
		\draw[->, line width=1pt] (-1,-0.5*2+0.7) node [font=\Large, above] {$\gamma_\mathrm{UT}$} -- (-0.2,-0.5*2+0.7);
		\draw[->, line width=1pt] (-1,-0.5*1+0.7) node [font=\Large, above] {$\beta_\mathrm{UT}$} -- (-0.2,-0.5*1+0.7);
		\draw[->, line width=1pt] (-1,-0.5*0+0.7) node [font=\Large, above] {$\alpha_\mathrm{UT}$} -- (-0.2,-0.5*0+0.7);
		\draw[->, line width=1pt] (-1,0.5*1+0.7) node [font=\Large, above] {$z_\mathrm{UT}$} -- (-0.2,0.5*1+0.7);
		\draw[->, line width=1pt] (-1,0.5*2+0.7) node [font=\Large, above] {$y_\mathrm{UT}$} -- (-0.2,0.5*2+0.7);
		\draw[->, line width=1pt] (-1,0.5*3+0.7) node [font=\Large, above] {$x_\mathrm{UT}$} -- (-0.2,0.5*3+0.7);

		\draw[->, line width=1pt, mycolor4] (8.15+2,-1) -- (9+2,-1) node [font=\Large, above right = -2pt and -17pt] {$P_{N_{P}|i}$};
		\draw[->, line width=1pt, mycolor4] (8.15+2,0.3) -- (9+2,0.3) node [font=\Large, above right = -2pt and -17pt] {$P_{2|i}$};
		\draw[->, line width=1pt, mycolor4] (8.15+2,1) -- (9+2,1) node [font=\Large, above right = -2pt and -17pt] {$P_{1|i}$};
		
		\node[align=left] at (1,-3.2) {\large Dense \\ Embedding};
		\node[align=left] at (3,-3.2) {\large Concatenate};
		\node[align=left] at (5,-3.2) {\large Dense};
		\node[align=left] at (7,-3.2) {\large Dense};
		\node[align=left] at (9,-3.2) {\large Dense};
		\end{tikzpicture}}}
	\caption{Multi-network structure of proposing candidate beam-panels by finding the proper AP beams and UT panels sequentially.}
	\label{Fig:NetStruc2}
\end{figure}
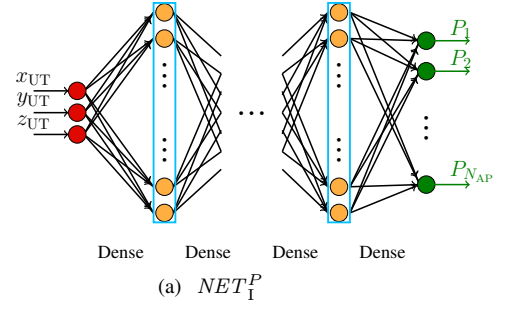
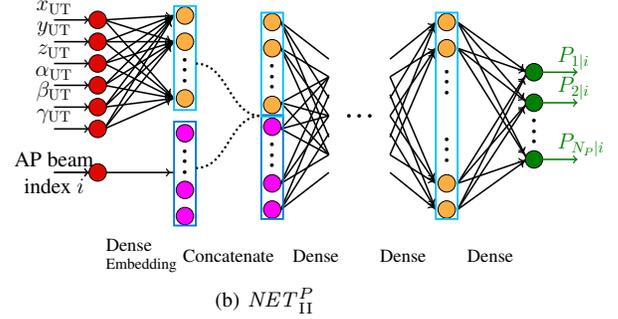
The multi-network panel selection (MN-PS) design uses two networks to propose pairs of AP beams and UT panels. 
Fig. \ref{Fig:NetStruc2} shows the networks $\mathit{NET}^P_{\mathrm{I}}$ and $\mathit{NET}^P_{\mathrm{II}}$ of this method. The role of $\mathit{NET}^P_{\mathrm{I}}$ is to sort AP beams based on the UT location. The outputs of $\mathit{NET}^P_{\mathrm{I}}$ are therefore estimates of $P_{i} = \mathbb{P} \left[ i = i^\star \right]$. $\mathit{NET}^P_{\mathrm{I}}$ has $4 n_h + (N_h-1) (n_h + 1) n_h + (n_h+1) N_\mathrm{AP}$ trainable parameters. $\mathit{NET}^P_{\mathrm{II}}$ includes an embedding layer that maps the AP beam index to a point in $\mathbb{R}^{n_h/2}$. The mapping is also a part of the learning process in training. As antenna panels are placed on different sides of the device, rotations of the device may change the optimal reception panel. Thus, both the UT location and orientation information are fed to $\mathit{NET}^P_{\mathrm{II}}$. $\mathit{NET}^P_{\mathrm{II}}$ includes $(7+N_\mathrm{AP}) n_h/2 + (N_h-1) (n_h + 1) n_h + (n_h+1) N_P$ trainable parameters.

We run $\mathit{NET}^P_{\mathrm{II}}$ multiple times with different indices of AP beams to get estimates of the optimality probabilities of each panel for the chosen AP beam, $P_{p|i} = \mathbb{P} \left[ p = p^\star | i = i^\star \right]$. The estimates of joint probabilities of AP beam $i$ and UT panel $p$ as the optimal choice  can be obtained as
\begin{equation}
    P_{i, p} = P_{p|i} P_{i}.
\end{equation}
A beam-panel candidate list can be made by selecting the combinations providing the highest probabilities. In this method, all the beams at the selected panels should be sensed by the UT. In case of using multiple RF chains at UT, in a time slot with a fixed beamforming vector $i^\star$ at AP, we can sense the environment simultaneously with $N_{RF}-1$ other panels that provide higher $P_{i^\star, p}$. MN-PS is robust to inaccuracy in orientation information, as slight changes in orientation may change the optimal UT beam, but the optimal panel is less likely to change. If UT has one RF chain per panel, $\mathit{NET}^P_{\mathrm{II}}$ is not needed, and the BA can be done using only UT location.

\subsection{Multi-Network Beam Selection Design}
In the multi-network beam selection (MN-BS) design, we follow the multi-network structure. The design of $\mathit{NET}^B_{\mathrm{I}}$ is exactly like $\mathit{NET}^P_{\mathrm{I}}$ in the UT panel selection method. The structure of $\mathit{NET}^B_{\mathrm{II}}$ is the same as $\mathit{NET}^P_{\mathrm{II}}$ except with $N_{UT}$ neurons at the output instead of $N_{P}$. $\mathit{NET}^B_{\mathrm{II}}$ provides estimates of conditional probabilities of UT beam $j$ as the optimal beam for a given AP beam, i.e.,
    \begin{equation}
        P_{j|i} = \mathbb{P} \left[ j = j^\star | i = i^\star \right].
    \end{equation}
The joint probabilities of transceivers beam pair $(i, j)$ as the optimal beam pair can be written as:
\begin{equation}
    P_{i, j} = P_{j|i} P_{i}.
\end{equation}
The candidate beam list $\mathcal{S}$ is made of beam-pairs with the highest estimated probabilities $P_{i, j}$ of being optimal. In case of having multiple RF chain at UT, we use the approach described in Section~\ref{SN} to sense with $N_{FR}$ panels simultaneously.

\begin{figure}[t]
	\centering
	\scalebox{0.12}{
		\includegraphics[trim={0cm 0cm 0cm 0cm},clip]{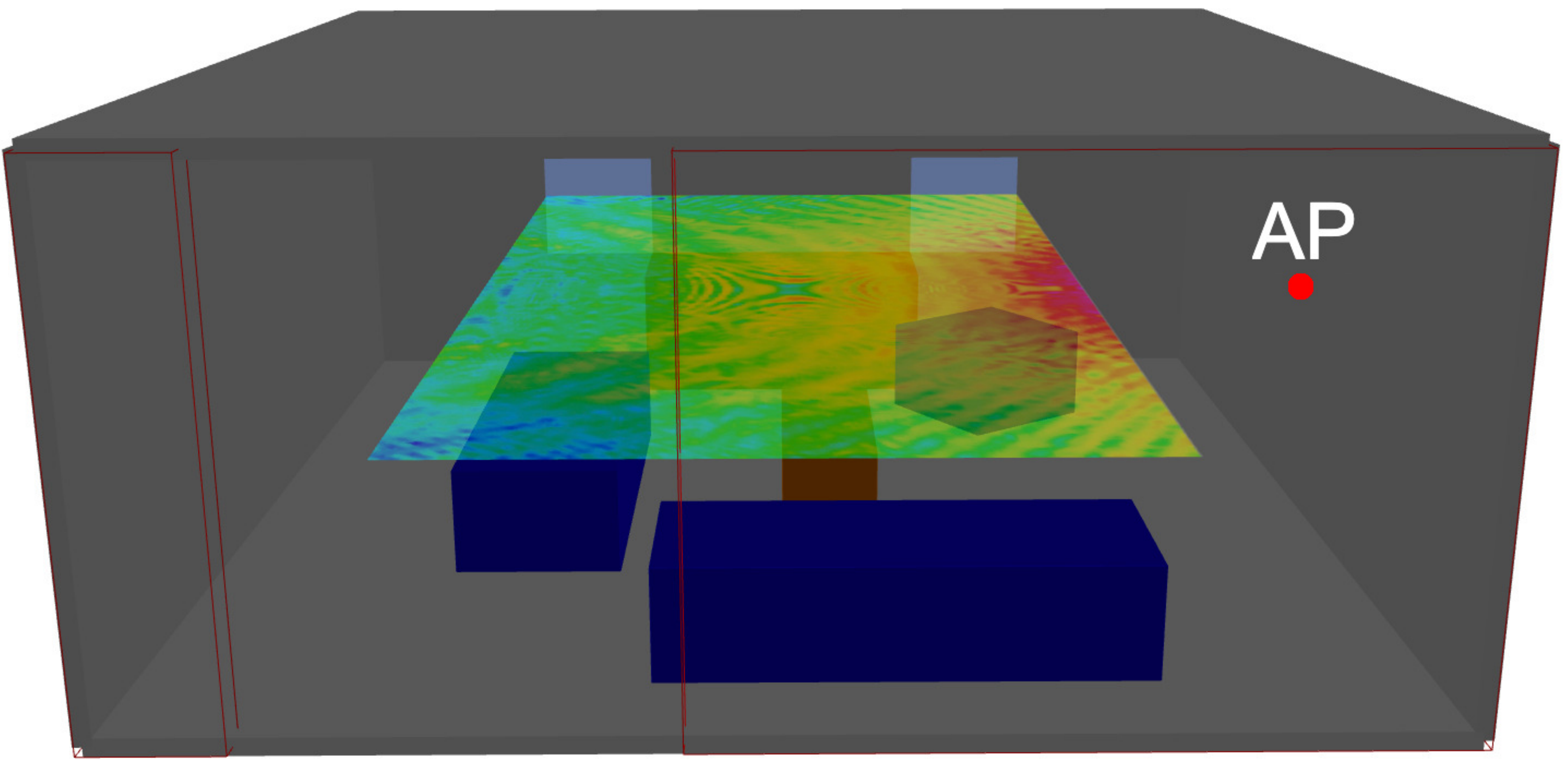}
	}
	\caption{The living room, an IEEE standard indoor scenario, in the ray-tracing simulation. The received LOS power at the user region is illustrated.}
	\label{Fig:LR}
\end{figure}
\section{Simulation Results}
In this study, we consider the living room proposed in the IEEE 802.11ad task group as an indoor environment with $7\times7\times3$ (m) dimension. The scenario and its propagation properties are described in detail in \cite{maltsev_channel_2010}. As shown in Fig. \ref{Fig:LR}, the AP is fixed and located in the center of one of the side walls. 
 The AP is made of a UPA panel with $\{1, 8, 8\}$ antenna elements. In this evaluation, we consider both designs for antenna placement shown in Fig. \ref{Fig:Designs}. Panels P$1$ to P$3$ have ULAs with $4$ antenna elements and UPAs with the configuration of $\{2, 2, 1\}$ antenna elements are used for P$4$ and P$5$. We use the antenna radiation pattern described in \cite{3gpp_study_2020} to simulate the antenna gain of a patch antenna. In addition, $P_t = 24~$dBm, and $\sigma_n^2 = -84~$dBm are used. For all the networks, we consider $N_h = 5$ and $n_h = 128$ and follow the training procedure explained in \cite{rezaie_location-_2020}. Thus, for the edge-face design, the SN, MN-PS, and MN-BS include $232K$, $146K$, and $148K$ trainable parameters, respectively.

\begin{figure}[t]
	\centering
	\subfloat[\label{Fig:AntFactor_a}]{%
       \scalebox{1}{\includegraphics[trim={3.5cm 0cm 8.2cm 1.05cm},clip,width=0.4\textwidth]{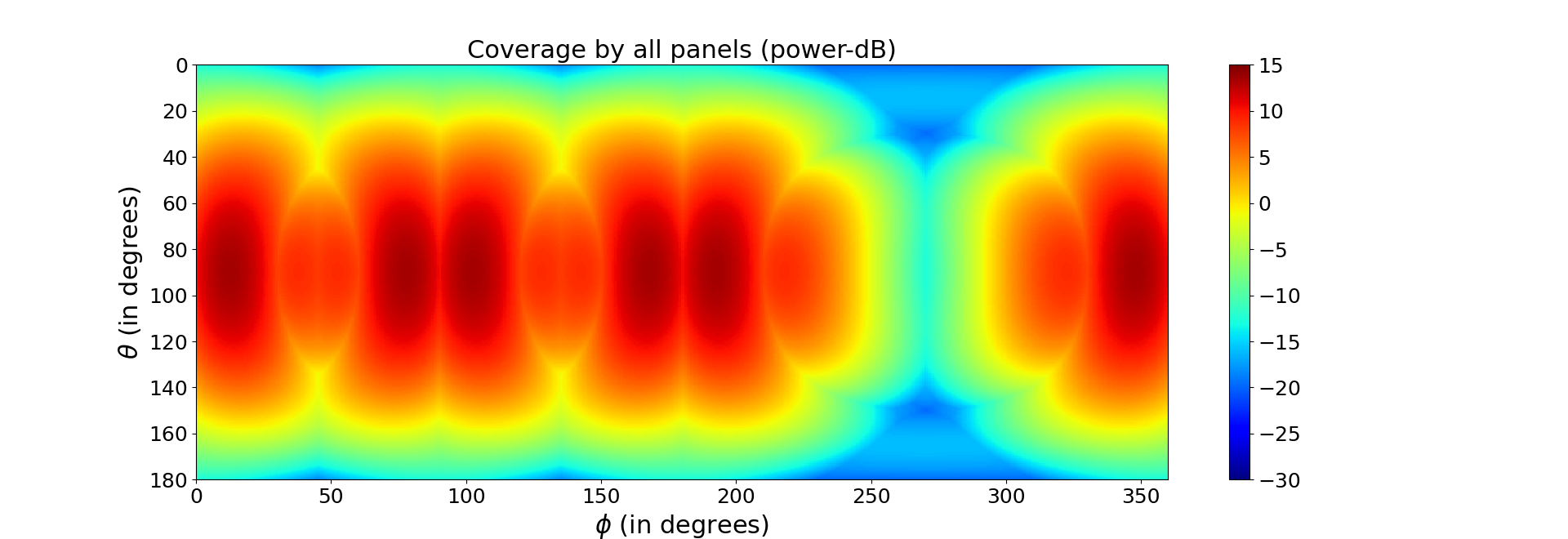}}}
    \\[0.05cm]
  \subfloat[\label{Fig:AntFactor_b}]{%
        \scalebox{1}{\includegraphics[trim={3.5cm 0cm 8.2cm 1.05cm},clip,width=0.4\textwidth]{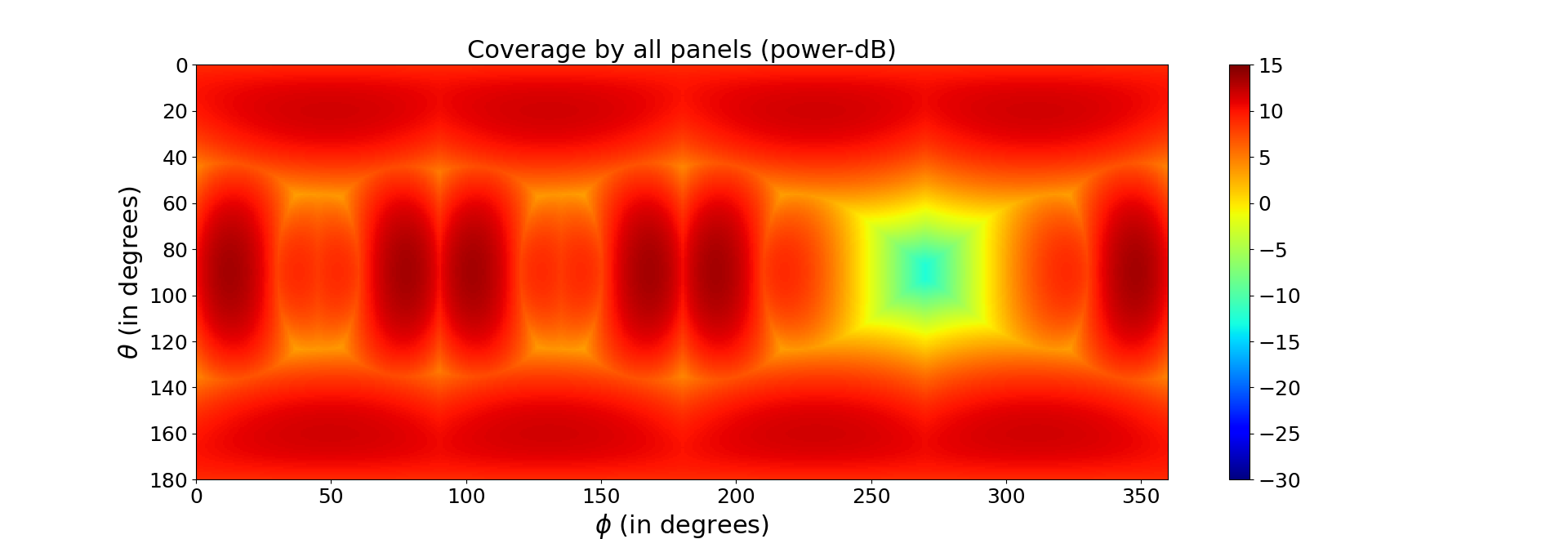}}}
	\caption{Antenna array factor for (a) Edge design (b) Edge-face design.}
	\label{Fig:AntFactor}
\end{figure}
Fig. \ref{Fig:AntFactor} shows the spherical coverage with all the panels for both considered designs. The edge design provides lower antenna gain in the directions perpendicular to the device face and back, and suffers from the lack of panel at the bottom. To evaluate the performance of DL-based BA methods, we consider two baselines: GIFP and HP-BS methods. The GIFP is a look-up table method and is described in detail in \cite{va_inverse_2018, rezaie_deep_2021-1}. In the HP-BS, the UT senses the environment with a wide beam for each panel while the AP transmits with a wide beam. Then, all  beams of the selected panel are used to  find the UT beam that provides the highest RSS. Later, the AP finds the best AP beam using a HBS algorithm \cite{rezaie_deep_2021-1}.

The UT position is uniformly drawn from the user grid, and the UT orientation is $50\%$ in portrait mode and $50\%$ in landscape mode, as defined in Section \ref{Sec:SysModel}. The line-of-sight (LOS) path is available in half of the realizations, while the other half is generated in non-LOS conditions. The AP and UT sense the environment with all possible combinations of beamforming vectors and combiners. We collect the RSS measurements in a dataset, besides each sample's UT location and orientation. To evaluate the performance of CI-based methods with different training dataset sizes, training datasets $\mathbb{D}^{\Xi}_{1}$ and $\mathbb{D}^{\Xi}_{2}$ respectively include  $56,000$ and $560$ training samples. We use a test dataset with $14,000$ samples for evaluation.

\subsection{Numerical Evaluation}

\begin{figure}[t]
	\centering
	\subfloat[ \label{Fig:EF56000_a}]{%
       \scalebox{1}{\includegraphics[trim={1.5cm 0.5cm 1cm 2cm},clip,width=0.4\textwidth]{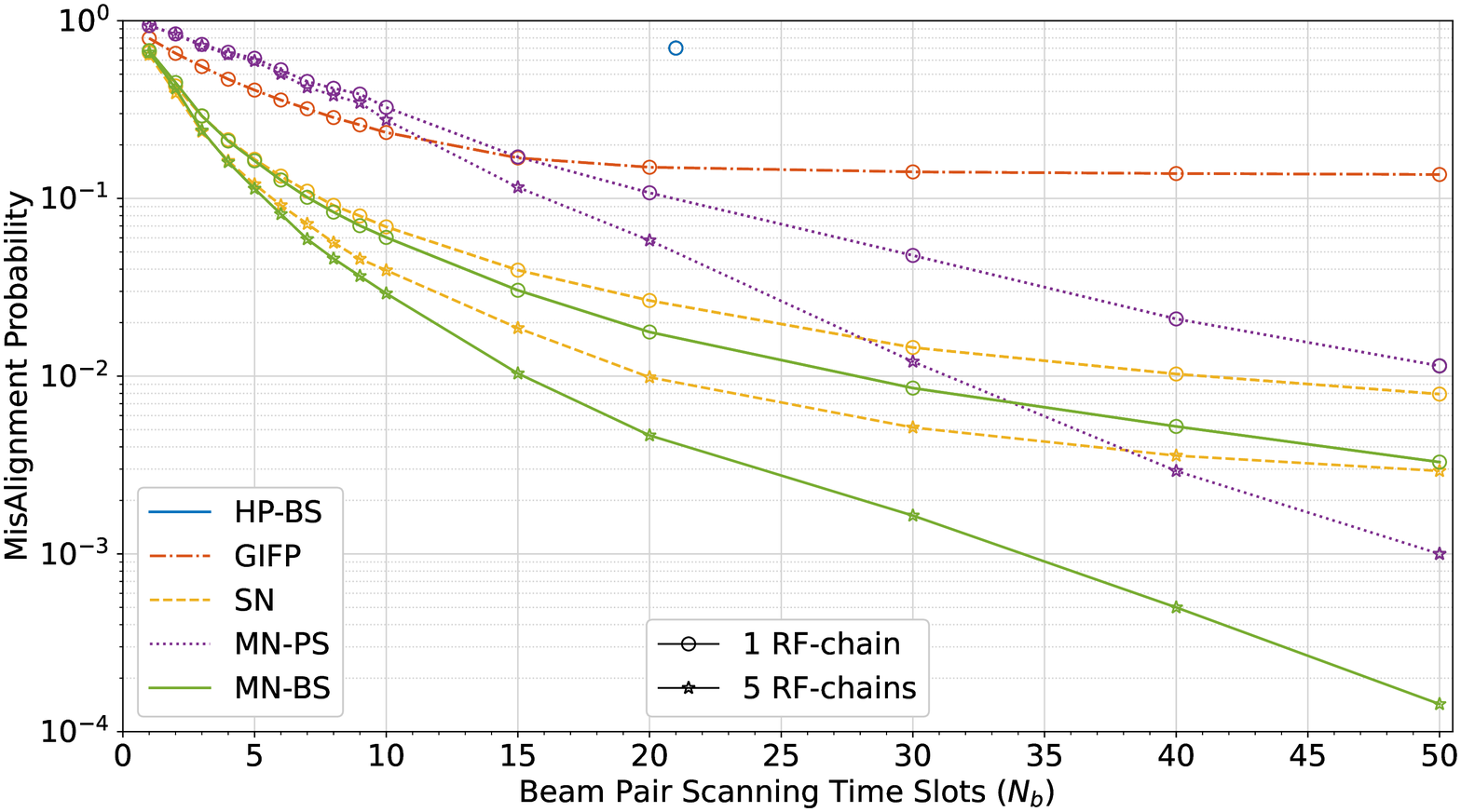}}}
    \\[0.05cm]
   \subfloat[\label{Fig:EF56000_b}]{%
        \scalebox{1}{\includegraphics[trim={1.5cm 0.5cm 1cm 2cm},clip,width=0.395\textwidth]{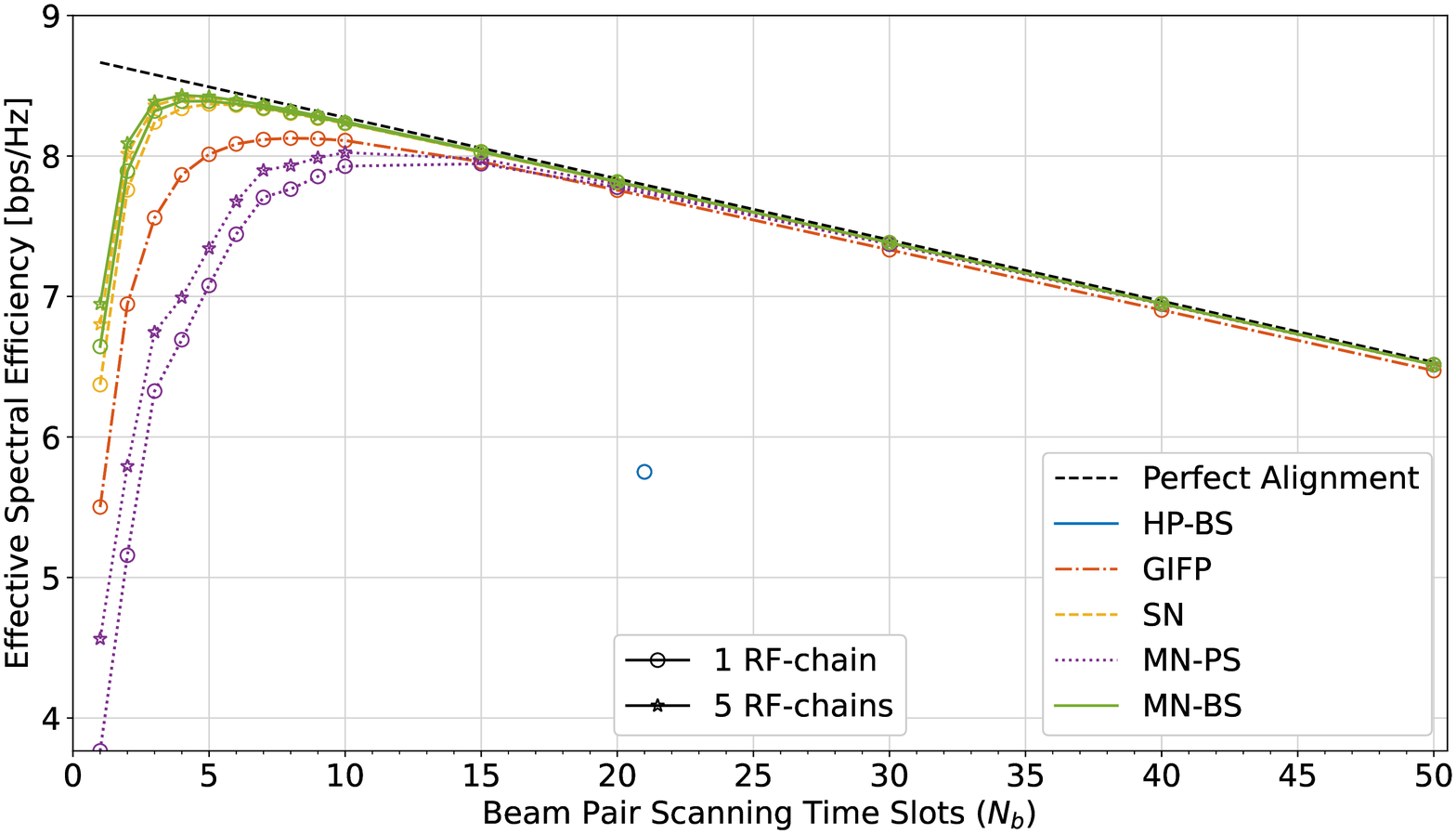}}}
	\caption{Misalignment probability and effective spectral efficiency for the edge-face design with $56,000$ training samples in $\mathbb{D}^{\Xi}_{1}$.}
	\label{Fig:EF56000}
\end{figure}
The performance\footnote{The effective spectral efficiency of a BA procedure using $N_b$ time slots is $\text{SE}_{\text{eff}} = \frac{T_{fr} - N_b T_s}{T_{fr}} \log_2 (1+ \text{SNR}_{\hat{i}, \hat{j}})$, where $\text{SNR}_{i, j}$ denote the SNR of $(i,j)$th beam pair. $T_{fr}=20$ms and $T_{s}=0.1$ms are used in this study \cite{rezaie_deep_2021-1}.} of the described BA methods using $\mathbb{D}^{\Xi}_{1}$ with $56,000$ training samples is shown in Fig. \ref{Fig:EF56000}. The MN-BS method has the highest accuracy, which shows the power of the multi-network design using the embedding layer. Also, we see almost no performance degradation in DL-based methods using only $1$ RF-chain instead of all $5$ RF-chains at the device. Thus, the UT can save power during the BA procedure by turning off all the panels except one of them. The MN-PS method senses all the beams at each panel, which leads to inefficiency compared to MN-BS. However, MN-PS with $5$ RF chains can perform well using only the UT location as a unique solution. CI-based BA methods using a large training dataset outperform the HB-PS method significantly.

\begin{figure}[t]
	\centering
	\subfloat[ \label{Fig:EF560_a}]{%
       \scalebox{1}{\includegraphics[trim={1.5cm 0.5cm 1cm 2cm},clip,width=0.395\textwidth]{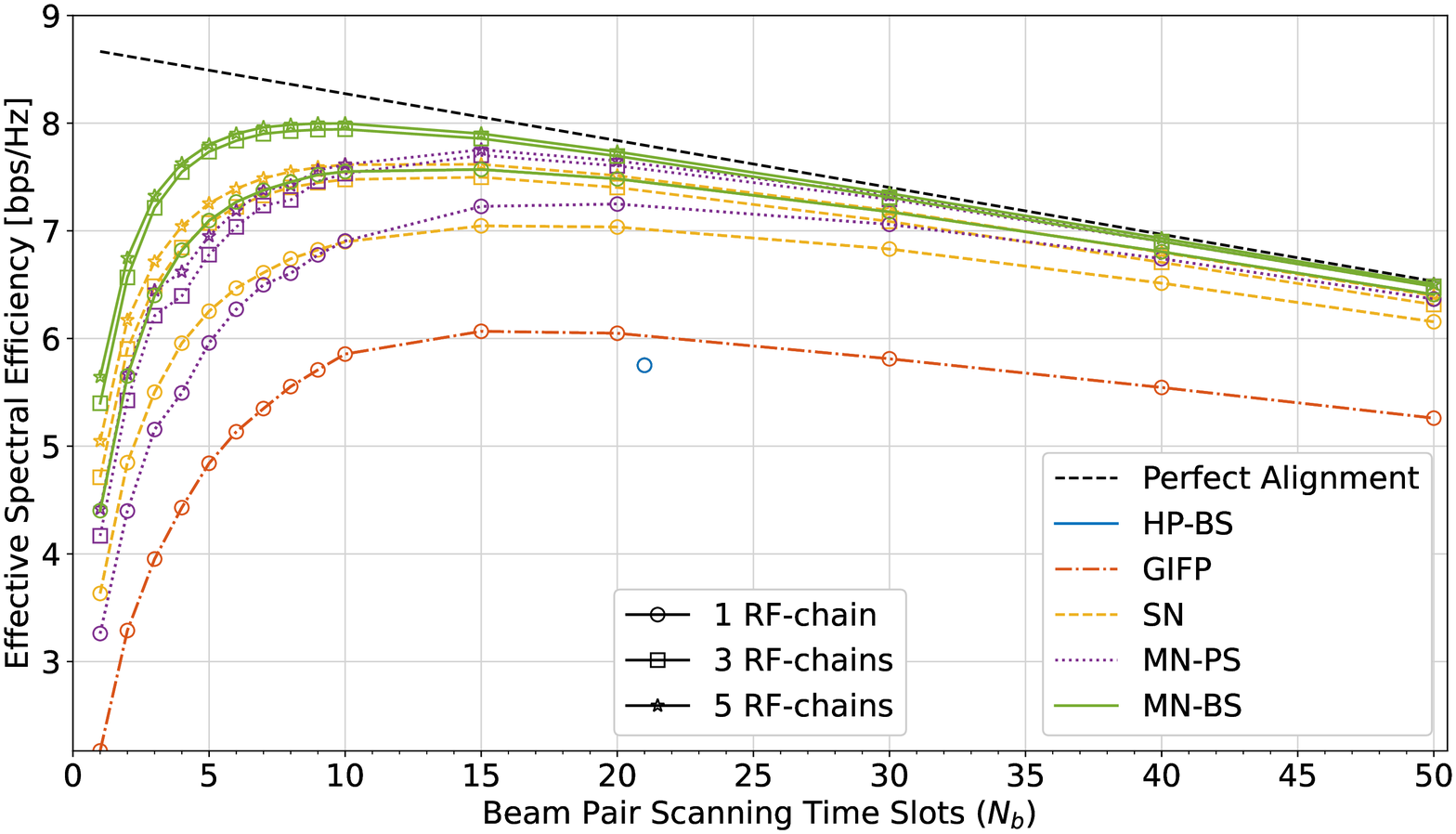}}}
    \\[0.05cm]
   \subfloat[\label{Fig:EF560_b}]{%
        \scalebox{1}{\includegraphics[trim={1.5cm 0.5cm 1cm 2cm},clip,width=0.395\textwidth]{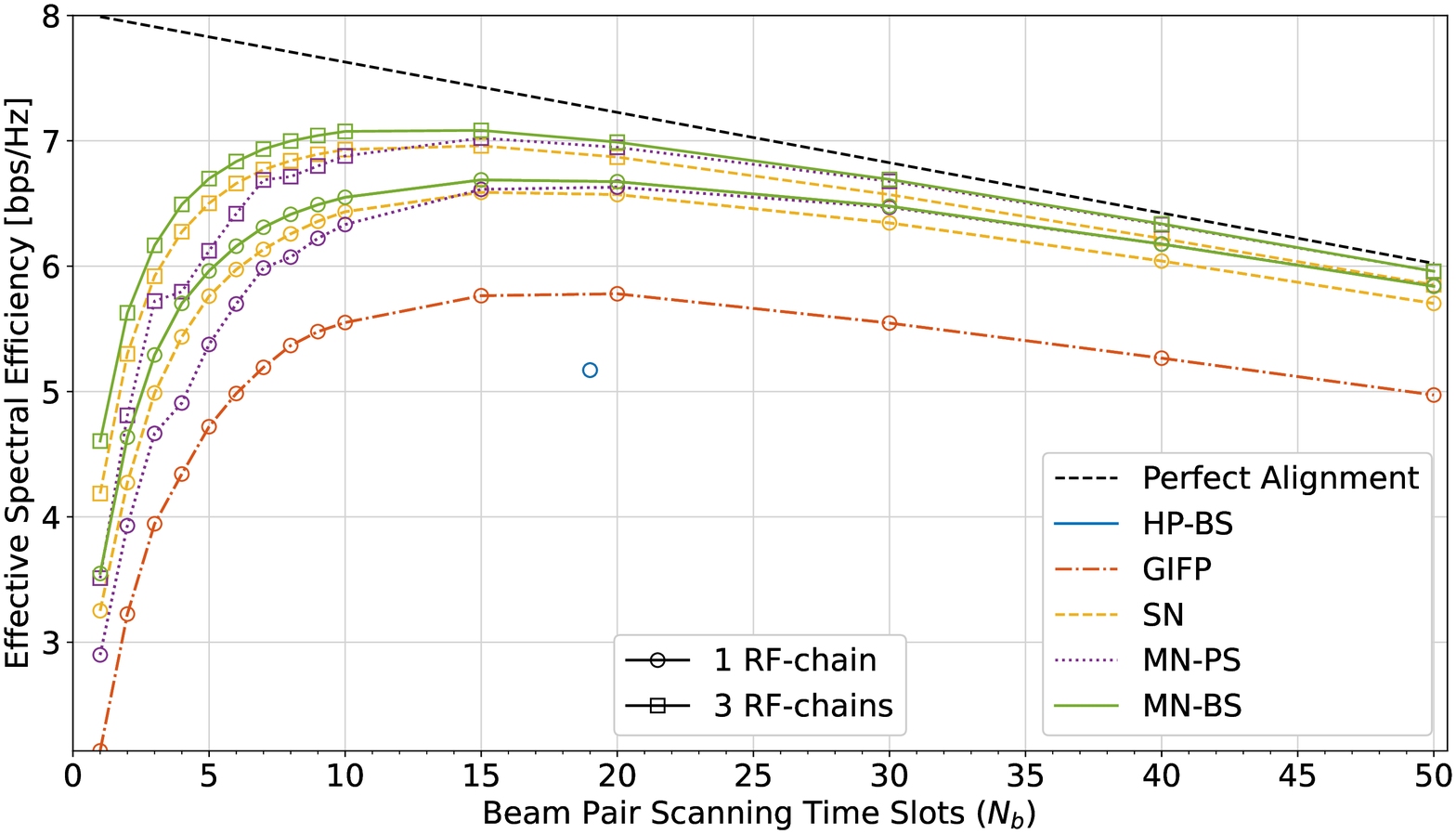}}}
	\caption{Performance of beam alignment methods with $560$ training samples in $\mathbb{D}^{\Xi}_{2}$ (a) edge-face design (b) edge design.}
	\label{Fig:EF560}
\end{figure}

Although transfer learning (TL) is an option for DL-based beam selection \cite{rezaie_deep_2021}, it is interesting to examine the performance of DL-based methods with small training datasets. Fig. \ref{Fig:EF560}\subref{Fig:EF560_a} shows the performance of the different BA methods using $\mathbb{D}^{\Xi}_{2}$ with $560$ training samples. Having fewer trainable parameters, the MN designs offer a significant gain compared to the SN structure. 
The edge design has advantages over the edge-face design in hardware complexity and cost with only a slight performance degradation, as shown in Fig. \ref{Fig:EF560}\subref{Fig:EF560_b}.

\section{Conclusions}
This work shows the usefulness of UT location and orientation in the initial BA of multi-panel devices in mmWave communications. Our results show that the deep learning-based methods offer excellent performance in proposing proper beam/panels which leads to power saving by turning off panels for a given UT coordinate and orientation. The numerical evaluations show that the multi-network designs provide the best performance in limited training dataset sizes due to having fewer trainable parameters than the single-network structure. The edge design can offer comparable performance to the edge-face design while reducing the cost and power at UT.
\bibliographystyle{IEEEtran} 
\bibliography{BeamAlignment_Ref} 

\end{document}